\documentclass[12pt,preprint]{aastex}

\shortauthors{Liu et al.}

\begin{document}

\title{A Pileup of Coronal Mass Ejections Produced the Largest Geomagnetic Storm in Two Decades} 

\author{Ying D. Liu\altaffilmark{1,2}, Huidong Hu\altaffilmark{1}, Xiaowei Zhao\altaffilmark{3,4}, Chong Chen\altaffilmark{5}, and Rui Wang\altaffilmark{1}}

\altaffiltext{1}{State Key Laboratory of Space Weather, National Space Science Center, 
Chinese Academy of Sciences, Beijing, China}

\altaffiltext{2}{University of Chinese Academy of Sciences, Beijing, China}

\altaffiltext{3}{National Satellite Meteorological Center, China Meteorological Administration, Beijing, China}

\altaffiltext{4}{School of Earth and Space Sciences, Peking University, Beijing, China}

\altaffiltext{5}{School of Microelectronics and Physics, Hunan University of Technology and Business, Changsha, China}

\begin{abstract}

The largest geomagnetic storm in two decades occurred in 2024 May with a minimum $D_{\rm st}$ of $-412$ nT. We examine its solar and interplanetary origins by combining multipoint imaging and in situ observations. The source active region, NOAA AR 13664, exhibited extraordinary activity and produced successive halo eruptions, which were responsible for two complex ejecta observed at the Earth. In situ measurements from STEREO A, which was $12.6^{\circ}$ apart, allow us to compare the ``geo-effectiveness" at the Earth and STEREO A. We obtain key findings concerning the formation of solar superstorms and how mesoscale variations of coronal mass ejections affect geo-effectiveness: (1) the 2024 May storm supports the hypothesis that solar superstorms are ``perfect storms" in nature, i.e., a combination of circumstances resulting in an event of an unusual magnitude; (2) the first complex ejecta, which caused the geomagnetic superstorm, shows considerable differences in the magnetic field and associated ``geo-effectiveness" between the Earth and STEREO A, despite a mesoscale separation; and (3) two contrasting cases of complex ejecta are found in terms of the geo-effectiveness at the Earth, which is largely due to different magnetic field configurations within the same active region.  

\end{abstract}


\section{Introduction}

In 2024 May the Sun exhibited substantial activity. A large complex active region resulted from the merging of NOAA AR 13664, which appeared on the east limb on April 30, and NOAA AR 13668, which emerged a few days later. Since the two active regions cannot be separated, we treat them as a single complex active region, referred to hereinafter as NOAA AR 13664. The complex active region disappeared on the west limb (i.e., rotated to the backside of the Sun) on May 13. As it rotated with the Sun from the east to west, it produced a series of M/X class solar flares and coronal mass ejections (CMEs). A prolonged geomagnetic storm
began on May 10 and reached a minimum $D_{\rm st}$ of $-412$ nT. The $D_{\rm st}$ index measures the intensity of geomagnetic storms: the more negative value, the more intense the storm. The only storm in the 21st century stronger than the current one is the 2003 November storm, which had a minimum $D_{\rm st}$ of $-422$ nT.   

The 2024 May storm can be classified as an extreme space weather event, i.e., a low-probability but high-consequence event otherwise called a solar superstorm \citep{report08, cannon13, riley18}. A well-known example of extreme space weather before the space era is the 1859 Carrington event \citep{carrington1859}, with a minimum $D_{\rm st}$ estimated to be about $-850$ nT \citep{siscoe06}. The most severe event of the space age is the 1989 March storm with a minimum $D_{\rm st}$ of $-589$ nT \citep{allen1989}. However, if the term ``solar superstorms" includes extreme solar wind disturbances that do not necessarily hit the Earth, then we have witnessed more cases, and extreme space weather may be more frequent than we imagine. In the last solar cycle (solar cycle 24), two solar superstorms occurred despite it being a historically weak cycle. The first was the 2012 July event \citep[e.g.,][]{baker13, russell13, liu14}, which was Carrington-class, and the second was the 2017 July event \citep{liu19}, which could rival the 1989 March storm. Although they did not hit the Earth, we infer that these two events would have produced geomagnetic superstorms if they had been Earth directed.    

In addition to creating a historic record, the 2024 May storm, with modern imaging and in situ observations from multiple vantage points, provides a significant opportunity to study the formation of solar superstorms. Important clues have already been learned from the 2012 July and 2017 July events. Based on the observations of the 2012 July event, \citet{liu14} suggest a ``perfect storm" scenario for the formation of an extreme event at 1 AU: in-transit interaction between successive eruptions to preserve the magnetic field, plus a preconditioning effect to prevent deceleration (i.e., one or several earlier eruptions clear the path). \citet{liu19} also performed a comparative study of the 2012 July and 2017 July events. They identify similar patterns in both cases: ``a long-lived eruptive nature of the source active region, a complex event composed of successive eruptions from the same active region, and in-transit interaction between the successive eruptions resulting in exceptionally strong ejecta magnetic fields at 1 AU." \citet{liu19} indicate that the concept of ``preconditioning" is a necessary condition for a Carrington-class storm to occur. They further propose a hypothesis that solar superstorms are essentially ``perfect storms," i.e., a combination of circumstances that produces an event of an unusual magnitude. It is worth mentioning that historical records of some extreme events support this hypothesis. For instance, revisits of the 1859 Carrington and 1989 March events with modern perspectives suggest that both resulted from successive CMEs from the same active regions \citep{green06, boteler19}. The recent 2022 September 5 CME, which is also considered as an extreme event, occurred from a complex active region with high productivity of flares and CMEs \citep{paouris23}; the authors also highlight the importance of preconditioning as a key factor for a high speed at 1 AU. It is intriguing to see how the 2024 May case will fit or challenge this hypothesis.   

Another merit of the 2024 May storm is that it allows us to investigate how the magnetic structure of CMEs varies across a mesoscale distance\footnote{There is no rigorous definition of ``mesoscale" for CMEs. In a discussion of multipoint measurements of CMEs, \cite{lugaz18} argue that the parameter space with radial separations of 0.005 - 0.05 AU and longitudinal separations of 1$^{\circ}$ - 12$^{\circ}$, which they call a mesoscale region, is left unexplored. Note that the size of a CME increases with distance from the Sun. The size near the Sun could be orders of magnitude smaller than that at a few AU. Here we tentatively define that the mesoscale for a CME ranges from MHD scale up to the radial width of the CME. At 1 AU, the average radial width of CMEs is 0.2 - 0.3 AU \citep[e.g.,][]{liu05, richardson10}, which translates to an angular size of about $14^{\circ}$. Specifically, the upper limit is $\sim$$14^{\circ}$ for the mesoscale of CMEs. However, our definition is still a subjective undertaking.}, and in particular, how the mesoscale variations affect geo-effectiveness. It has in situ measurements at both the Earth and STEREO A (see below), which were separated in longitude by about $12.6^{\circ}$ at that time. The magnetic fields of the CMEs at the Earth and STEREO A could be quite different, even if their flux-rope orientations are the same at the two points. This may result in different $D_{\rm st}$ values (i.e., different ``geo-effectiveness") at the two points. In the STEREO era, the first CME that was measured in situ by multiple spacecraft at a mesoscale separation was the 2007 May event \citep{liu08, kilpua09, mostl09}, during which the Earth and STEREO B were about $3^{\circ}$ apart. \citet{liu08} examined the magnetic field measurements of the CME at the Earth and STEREO B. They find that, while the measurements at the two points are consistent with a flux-rope geometry, they show a considerable difference in the magnetic field (both strength and components) even for a $3^{\circ}$ separation. Similar results have been obtained with multipoint measurements of a few degrees apart \citep[e.g.,][]{lugaz18, regnault24, palmerio24}. \citet{regnault24} warn that, if in situ measurements at a mesoscale separation are used to predict the southward field component at the Earth, a large error may result. The current case is a superstorm, and we have the rare opportunity to compare the ``geo-effectiveness" of a superstorm at two points separated by a mesoscale distance. 

While this Letter may serve as a description of the timeline leading up to the geomagnetic superstorm, it also attempts to: (1) provide a timely analysis of the solar and interplanetary origins of the 2024 May geomagnetic superstorm; (2) examine if the 2024 May storm conforms to the hypothesis of \citet{liu19} that solar superstorms are ``perfect storms" in nature; and (3) investigate how the mesoscale variations of CME magnetic fields affect geo-effectiveness. We first identify full halo eruptions that are the most likely candidates impacting the Earth. We then perform reconstruction of the CMEs to determine their parameters near the Sun. Time-elongation maps from wide-angle imaging observations are constructed in order to see their propagation behaviors in the Sun-Earth space, which may help enhance geo-effectiveness. We also examine the solar wind signatures at 1 AU and their connections with the geomagnetic storm. Modeling of the $D_{\rm st}$ index is carried out, which allows us to compare the ``geo-effectiveness" between two different points. We describe multipoint imaging and in situ observations in Section~2, and conclude in Section~3.  

\section{Observations and Results} 

Figure~1 shows the positions of the spacecraft in the ecliptic plane as well as the propagation directions of some full halo CMEs. STEREO A was about 0.96 AU from the Sun and about $12.6^{\circ}$ west of the Earth on May 10. The Parker Solar Probe (PSP) was near its aphelion, i.e., about 0.74 AU from the Sun and around $95.4^{\circ}$ west of the Earth. Solar Orbiter (SolO) was located at about 0.69 AU and about $167.3^{\circ}$ west of the Earth. At L1 upstream of the Earth we have imaging observations from SOHO and in situ measurements from Wind, while STEREO A provides both imaging and in situ observations. The longitudinal separation between the Earth and STEREO A at almost the same distance is appropriate for an investigation of mesoscale variations of CME magnetic structures and associated ``geo-effectiveness."   

\subsection{White-light Imaging Observations}

We first identify full halo CMEs from LASCO imaging observations aboard SOHO. These are the most likely candidates that could have impacted the Earth. We start from May 1 when the active region appeared on the east limb, but all full halo CMEs occurred starting from May 8. Table~1 provides the information about the full halo eruptions, including the associated flares and source locations on the Sun. They are successive eruptions, all from NOAA AR 13664. The source location is determined from the brightest flare kernel on the Sun. The CME parameters are estimated with a graduated cylindrical shell (GCS) technique, which assumes a rope-like morphology for CMEs with two ends anchored at the Sun \citep{thernisien06, thernisien09}. Both views from SOHO and STEREO A are used in the modeling. Note that the CMEs also appeared as full halo events for STEREO A, since it was only $12.6^{\circ}$ apart from the Earth. For each event, the flux-rope tilt angle, aspect ratio and half angle are assumed to be fixed in the associated time series of GCS modeling; the propagation longitude and latitude are allowed to vary, but we observe little variations in these two parameters. The propagation longitude and latitude given in Table~1 are average values from the time series of the modeling. The CME propagation directions are generally consistent with the source locations on the Sun, although deviations are observed. For each CME, the velocity is derived from a linear fit of the leading-front distances in GCS reconstruction; projection effects are already removed. All the velocities in Table~1 are below 2000 km s$^{-1}$, so the CMEs are not extreme in terms of speeds. The flux-rope tilt angles listed in Table~1 may have large uncertainties, as the CMEs are full halos for both spacecraft. 

The turbulent corona and inner heliosphere during the time period can be seen from the time-elongation maps in Figure~2, which are produced by stacking the running-difference images within a slit along the ecliptic plane \citep[e.g.,][]{sheeley08, davies09, liu10}. Also shown is the GOES X-ray flux (Figure~2 top), so we can associate the CMEs with the flares (EUV images are also examined for this association). Many CMEs occurred, but our focus is the full halo eruptions, which are revealed by the maps from both SOHO and STEREO A. Given the clustering of the CMEs from the same active region, some of them may interact during their propagation in the corona and inner heliosphere. We observe intersections of some tracks in HI1 and HI2 of STEREO A (Figure~2 bottom), which indicate CME-CME interactions. In particular, the CMEs from May 8 (at least) seem to merge into a large, bright front in HI2. According to our experience of connecting wide-angle imaging observations with in situ signatures, the merging of multiple tracks into a bright front in time-elongation maps implies in situ signatures of a forward shock followed by a complex ejecta \citep[see][]{liu12, liu20}. This is also what we expect at the Earth (which was at an elongation of about $97.9^{\circ}$) in the current case.

Comparing the tracks to the observed shock arrival times at the Earth is helpful to establish the connections between the CMEs and the in situ signatures observed at the Earth. The CME parameters from GCS modeling also provide some clues. We observe an increasing trend in the CME velocities for CMEs 1 - 4 and another increasing trend for CMEs 5 - 7 (see Figure~1 and Table~1). The speed distributions, together with the timing, suggest that CMEs 1 - 4 would pile up and interact, and that CMEs 5 - 7 would do the same. If this is true, we may expect two complex ejecta at the Earth, one from the merging of CMEs 1 - 4 and the other from the merging of CMEs 5 - 7. CME 8 is not relevant, as its direction is far too westward and its launch time is far too late (compared to the occurrence of the geomagnetic storm). Note that the average direction of CMEs 1 - 4 is more towards STEREO A than towards the Earth (see Figure~1), so we may anticipate a more head-on collision with STEREO A.          
 
\subsection{In Situ Measurements at Wind} 

The in situ measurements at Wind are shown in Figure~3. We see complex magnetic field and plasma signatures. In general, two complex ejecta can be identified from the measurements. As defined by \citet{burlaga01, burlaga02}, complex ejecta result from interactions between successive CMEs and usually do not have well ordered magnetic fields at 1 AU. Complex ejecta can be very geo-effective because of their prolonged durations and enhanced magnetic fields \citep[e.g.,][]{farrugia04, zhang07, lugaz14, liu15, liu20, mishra15}. Our identification of the complex ejecta is based on a combination of the depressed proton temperature, low proton $\beta$, and magnetic field signatures (such as enhancement, smoothness and indication of rotation), but this is a subjective undertaking. The second complex ejecta resembles a typical ICME (interplanetary counterpart of a CME), given the declining speed profile and the depressed temperature throughout its interval compared with the expected temperature. The magnetic field inside it, however, shows multiple polarities and the presence of a shock-like structure (around 23:21 UT on May 13), so we consider it as a complex ejecta. Each complex ejecta is associated with a forward shock, which passed Wind at 16:37 UT on May 10 and 09:09 UT on May 12, respectively. The second shock is propagating into the first complex ejecta, compressing the plasma and magnetic field. There is another shock passage at 19:02 UT on May 15 (not shown here), which was likely produced by CME 8 from May 13. The shock driven by the May 13 eruption must have a large longitudinal width (see Figure~1). 

The observations of two complex ejecta at 1 AU are consistent with our expectation in Section~2.1. There are multiple dips in the temperature profile inside the first complex ejecta (Figure~3c), indicative of multiple CMEs. The speed profile is not declining monotonically across the interval but shows four major bumps (Figure~3b), so the merging of the CMEs is still in process at 1 AU. The velocity is not particularly high, with a peak value of about 1000 km s$^{-1}$ within the complex ejecta. We suggest that the four major bumps correspond to CMEs 1 - 4 (see Figure~1 and Table~1), but other smaller events (not listed in Table~1) may also have contributed to the formation of the first complex ejecta. The second complex ejecta is presumably formed from the merging of CMEs 5 - 7, as mentioned earlier. Again, other smaller CMEs (not listed in Table~1) may also have contributed to its formation. 

Of particular interest is the extremely enhanced magnetic field inside the first complex ejecta. The magnetic field strength is as high as about 72 nT and the peak southward field component is about 59 nT, both of which are unusually large values at 1 AU. As discussed earlier, this large magnetic field must have been produced by the interactions between the successive CMEs during their transit to 1 AU, which amplify the magnetic field and help maintain a strong field. The enhanced southward field mainly takes place inside the first complex ejecta but not inside the second one, so we can contrast two cases of complex ejecta in terms of the magnetic field and geo-effectiveness. \citet{wang24} examine the active region magnetic field and suggest that CMEs 1 - 4 and CMEs 5 - 7 erupted from two different groups of polarity inversion lines (PILs): the first group of PILs, from which CMEs 1 - 4 occurred, is associated with a field distribution that implies strong southward fields; the opposite is true for the second group of PILs, from which CMEs 5 - 7 erupted. Readers are directed to \citet{wang24} for details. Also note that CMEs 5 - 7 are not as clustered as CMEs 1 - 4 (see Figure~2 bottom). Combined together, these may explain the much smaller magnetic field strength and much weaker southward field inside the second complex ejecta, although it is also a result of CME-CME interactions.     

We notice a shift in the flux-rope tilt angles of CMEs 1 - 7 from about $-60^{\circ}$ to about $60^{\circ}$ (see Table~1). This shift seems consistent with the alterations in the local $B_{\rm N}$ component from being predominantly southward inside the first complex ejecta to being predominantly northward within the second complex ejecta. It also appears to agree with the two different configurations of the active region magnetic fields mentioned above. While this shift is intriguing, we warn that the flux-rope tilt angles may have large uncertainties. They are derived from white-light images, and the CMEs are full halos for both the Earth and STEREO A.

The geomagnetic superstorm with a minimum $D_{\rm st}$ of $-412$ nT was caused by the first complex ejecta with its persistent, enhanced southward magnetic field (Figure~3i). It shows a prolonged recovery phase, during which the second complex ejecta only produced slight dips in the $D_{\rm st}$ profile. We determine the $D_{\rm st}$ index from the solar wind parameters using two empirical formulas \citep{burton75, om00}. The two schemes assume different decays for the terrestrial ring current, which result in different $D_{\rm st}$ profiles. According to our experience, the \citet{burton75} model tends to overestimate, while the \citet{om00} formula tends to underestimate, the $D_{\rm st}$ index. We therefore average their results. The resulting $D_{\rm st}$ profile is similar to the measured one. It has a minimum value of about $-378$ nT, only about 8\% smaller than the actual minimum. The similarity between the estimated and measured $D_{\rm st}$ index at the Earth can be considered as a ``calibration," which enables the application of our approach to the solar wind measurements at STEREO A.          

\subsection{In Situ Measurements at STEREO A} 

The in situ measurements at STEREO A are displayed in Figure~4. The plasma parameters are from level 2 preliminary Maxwellian fits and have many data gaps. Note that we have shifted the data at Wind forward in time by about 2.6 hr in order to align the first shock at each spacecraft. Specifically, the shock associated with the first complex ejecta arrived at STEREO A at 14:03 UT on May 10, which is earlier by about 2.6 hr. The magnetic field profile is similar to that at Wind, but the field strength at STEREO A is generally stronger (Figure~4d). The solar wind velocity also seems higher at STEREO A (Figure~4b), although the data gaps bring some uncertainties. The stronger field strength and earlier arrival at STEREO A agree with a more head-on collision with STEREO A (see Figure~1), as we expected from coronagraph imaging observations. The data (particularly the magnetic field) indicate two other shock passages at STEREO A: 07:32 UT on May 12 and 00:05 UT on May 15 (with the latter corresponding to the May 13 CME). Both of them are also earlier than their counterparts at Wind. 

In addition to the field strength, we also observe considerable differences in the magnetic field components between Wind and STEREO A (Figure~4f - h), despite a longitudinal separation of only $12.6^{\circ}$. Note that radial variations are minimized since the two spacecraft are at similar distances from the Sun. These differences would affect geo-effectiveness associated with the magnetic structure. We evaluate the $D_{\rm st}$ index using the same approach as we have done on the solar wind measurements at the Earth. We feed the models with the magnetic field measurements at STEREO A. Since the plasma data at STEREO A have a lot of data gaps, we take the shifted plasma data at Wind as input. This gives a lower limit for the ``geo-effectiveness," as the solar wind velocity should be higher for a more head-on impact. The resulting $D_{\rm st}$ profile shows multiple dips (Figure~4i), as we have seen at the Earth. It has a minimum value of about $-494$ nT, which is much larger than its counterpart at the Earth (i.e., $-378$ nT). Therefore, the geomagnetic storm would have been much stronger, if the portion of the solar wind disturbance measured at STEREO A had hit the Earth. Clearly, a mesoscale separation can result in significant differences in both the magnetic field and associated geo-effectiveness. It is worth mentioning that here we look at a complex ejecta. The scenario for a single CME may (or may not) be different.  

Note that the result of a much stronger storm at STEREO A relies on the assumption about the solar wind velocity at STEREO A. The southward magnetic field component and the solar wind velocity, in order of importance, are two key elements controlling the intensity of geomagnetic storms. Measurements at STEREO A indicate a generally stronger southward field component (Figure~4h), but the velocity data are largely missing. Our approach using the shifted solar wind speed at Wind may carry a considerable uncertainty. However, the velocity measurements at STEREO A, although sparse, suggest that the speed at STEREO A is at least comparable to that at Wind (Figure~4b). Therefore, the conclusion of a stronger storm at STEREO A is valid, and this is consistent with a more head-on impact with STEREO A.

\section{Conclusions}

We have examined the solar and interplanetary sources of the 2024 May geomagnetic superstorm, the largest one in two decades. Key findings are revealed concerning the formation of solar superstorms and how CME mesoscale variations affect geo-effectiveness. We summarize the results as follows.  

(1) The 2024 May storm supports the hypothesis of \citet{liu19} that solar superstorms are ``perfect storms" in nature. The active region NOAA AR 13664 kept erupting for a prolonged time period. Among the series of eruptions from the active region, four halo CMEs (CMEs 1 - 4) on May 8 - 9 interacted during their transit to 1 AU, which was responsible for the unusually large magnetic field inside the first complex ejecta observed at the Earth. This is essentially a ``perfect storm" scenario, specifically, a combination of circumstances resulting in an event of an unusual magnitude. It is also exactly what \citet{liu19} find in their comparative study of the 2012 July and 2017 July events: ``a long-lived eruptive nature of the source active region, a complex event composed of successive eruptions from the same active region, and in-transit interaction between the successive eruptions resulting in exceptionally strong ejecta magnetic fields at 1 AU." Note that none of the present CMEs is extreme in terms of their velocity near the Sun. These results strengthen the idea that extreme events generally are not simple and can involve reinforcing factors. As in \citet{liu14, liu19}, the results also point out that extreme events are not as rare as we imagine. 

(2) The complex ejecta, which caused the geomagnetic storm, shows considerable differences in the magnetic field and associated ``geo-effectiveness" between the Earth and STEREO A, despite a longitudinal separation of only $12.6^{\circ}$. An earlier arrival and stronger field are observed at STEREO A, which is consistent with a more head-on collision with STEREO A as expected from coronagraph imaging observations. We evaluate the $D_{\rm st}$ index using the same approach for the two points, which allows us to compare the ``geo-effectiveness" at the Earth and STEREO A. The simulated $D_{\rm st}$ profile is similar to the measured one at the Earth, with a slightly underestimated minimum ($-378$ nT from the model versus the actual $-412$ nT). At STEREO A we obtain a lower limit for the ``geo-effectiveness" with a minimum $D_{\rm st}$ of $-494$ nT. The geomagnetic storm would be much stronger (compared with the counterpart $-378$ nT at the Earth), if the Earth were at the position of STEREO A. Therefore, a mesoscale separation can result in significant differences in CME magnetic field and associated geo-effectiveness.  

(3) We find two contrasting cases of complex ejecta in terms of their geo-effectiveness. The geomagnetic superstorm was caused by the first complex ejecta resulting from the merging of CMEs 1 - 4, due to its persistent, enhanced southward magnetic field. The second complex ejecta observed at the Earth, which was likely formed from the merging of CMEs 5 - 7, is not associated with a strong southward field. It only produced slight dips in the $D_{\rm st}$ profile during the recovery phase of the geomagnetic storm. \citet{wang24} indicate two different distributions of the active region magnetic field around the erupting PILs: one implies strong southward fields, but the other does not. Also, CMEs 5 - 7 are not as clustered as CMEs 1 - 4. These may explain their distinct geo-effectiveness, although both were a result of CME-CME interactions.  

\acknowledgments 
The research was supported by the Strategic Priority Research Program of the Chinese Academy of Sciences (No. XDB0560000), NSFC (under grants 42274201, 42204176, 12073032 and 42150105), and National Key R\&D Program of China (No. 2021YFA0718600). We acknowledge the use of data from Wind, STEREO and SOHO and the $D_{\rm st}$ index from WDC in Kyoto. Readers are directed to \citet{kaiser08} for the STEREO mission, \citet{domingo95} for the SOHO mission, \citet{ogilvie95} for Wind/SWE, and \citet{lepping95} for Wind/MFI.

\clearpage

\begin{figure}
\epsscale{0.7} \plotone{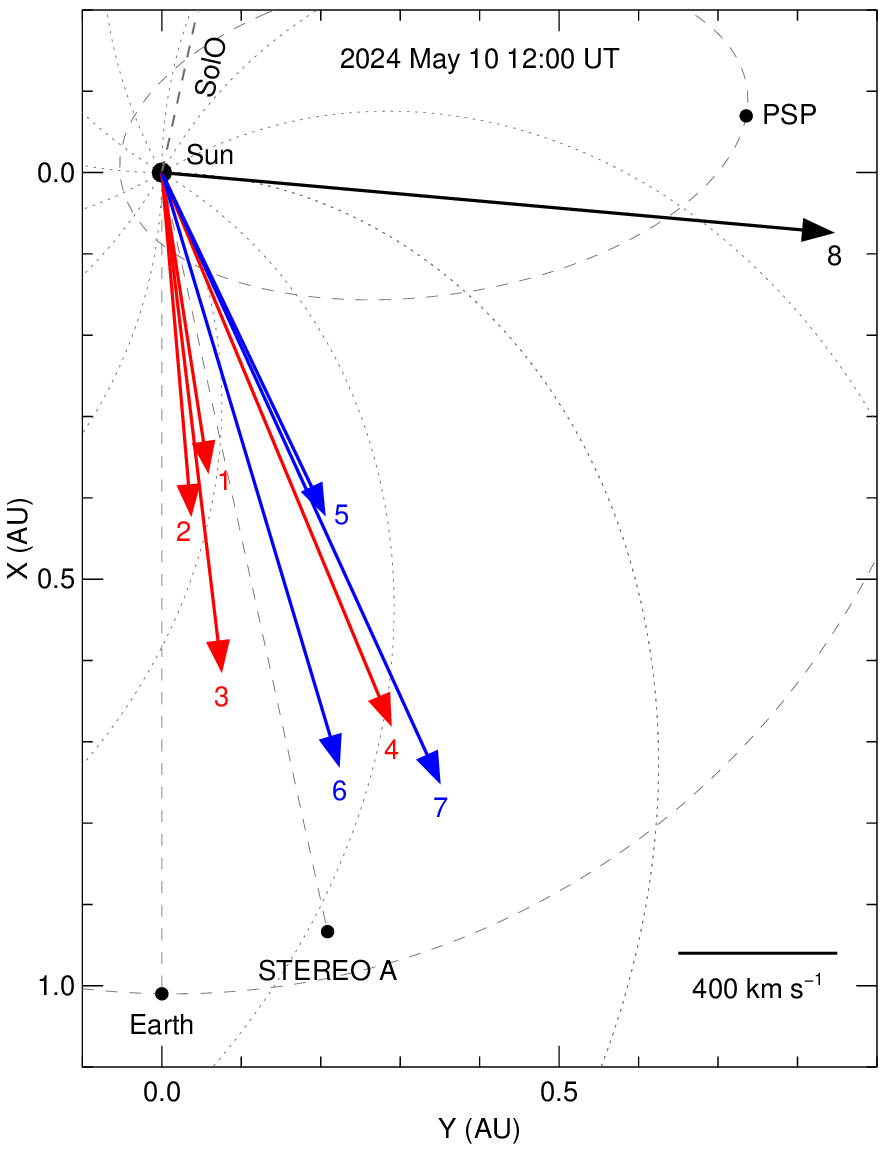} 
\caption{Positions of the spacecraft in the ecliptic plane at 12:00 UT on 2024 May 10. The dashed lines indicate the longitudes of the Earth (upstream of which Wind and SOHO are located), STEREO A and SolO, respectively. The gray dashed curves are the orbits of the Earth and PSP. The dotted lines show Parker spiral magnetic fields created with a solar wind speed of 450 km s$^{-1}$. The arrows mark the propagation directions of the full halo CMEs from May 8 - 13 obtained through CME reconstruction, with the lengths of the arrows indicating CME velocities near the Sun. Red (blue) arrows represent CMEs that likely contributed to the first (second) complex ejecta at the Earth. The numbers label the CMEs as in Table~1.}
\end{figure}

\clearpage

\begin{deluxetable}{lcccccccc}
\tabletypesize{\scriptsize}
\tablecaption{Information about the Full Halo CMEs Impacting the Earth}
\tablewidth{0pt}
\tablehead{\colhead{No.} & \colhead{Flare class} & \colhead{Flare peak time} & \colhead{Source location} 
& \colhead{CME direction} & \colhead{CME velocity} & \colhead{Tilt angle} & \colhead{Aspect ratio} & \colhead{Half angle} \\
  &  & (UT) &   &   & (km s$^{-1}$)  & & }
\startdata
1 & X1.0  & May 8, 05:09  & S17$^{\circ}$W09$^{\circ}$ & S14$^{\circ}$W09$^{\circ}$ & 750   &  $-63^{\circ}$  & 0.96 & $19^{\circ}$ \\
2 & M8.7 & May 8, 12:03   & S16$^{\circ}$W10$^{\circ}$ & S10$^{\circ}$W05$^{\circ}$ & 850   &  $-55^{\circ}$ & 0.95 & $15^{\circ}$ \\
3 & X1.0  & May 8, 21:40   & S18$^{\circ}$W18$^{\circ}$ & S16$^{\circ}$W07$^{\circ}$ & 1240 &  $-10^{\circ}$ & 0.72 & $9^{\circ}$  \\
4 & X2.2  & May 9, 09:13   & S18$^{\circ}$W23$^{\circ}$ & S12$^{\circ}$W23$^{\circ}$ & 1480 &  $-25^{\circ}$ & 0.84 & $8^{\circ}$ \\
\hline
5 & X1.1  & May 9, 17:44   & S15$^{\circ}$W28$^{\circ}$ & S06$^{\circ}$W26$^{\circ}$ & 940   &  $-3^{\circ}$   & 0.92 & $10^{\circ}$ \\
6 & X3.9  & May 10, 06:54 & S14$^{\circ}$W34$^{\circ}$ & S10$^{\circ}$W17$^{\circ}$ & 1530 &  $52^{\circ}$  & 0.62 & $6^{\circ}$ \\
7 & X5.8  & May 11, 01:23 & S15$^{\circ}$W44$^{\circ}$ & S05$^{\circ}$W25$^{\circ}$ & 1660 &  $58^{\circ}$  & 0.92 & $36^{\circ}$ \\
\hline
8 & M6.6 & May 13, 09:44 & S23$^{\circ}$W80$^{\circ}$ & S36$^{\circ}$W85$^{\circ}$ & 1700 &  $90^{\circ}$  & 0.70 & $25^{\circ}$ \\
\enddata
\end{deluxetable}

\clearpage

\begin{figure}
\epsscale{0.9} \plotone{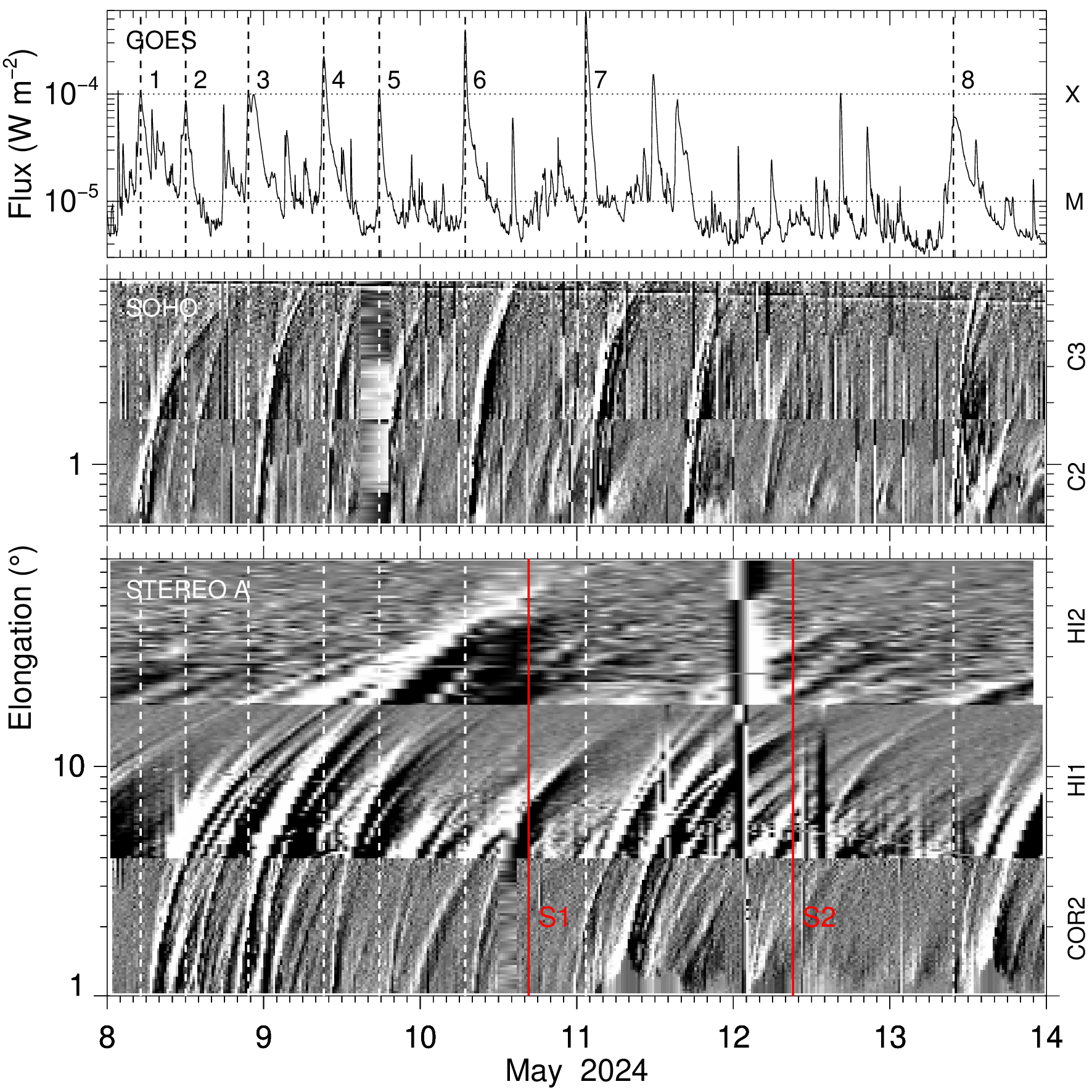} 
\caption{Full halo eruptions on May 8 - 13. Top: GOES X-ray flux at 1 - 8 \AA. Middle: Time-elongation maps from LASCO of SOHO (at a position angle of $90^{\circ}$ measured clockwise from the solar north). Bottom: Time-elongation maps from SECCHI of STEREO A (at a position angle of $270^{\circ}$ clockwise from the solar north). The vertical dashed lines indicate the peak times of the associated flares, with the numbers labelling the eruptions as in Table~1. The red vertical lines mark the observed arrival times of the shocks at the Earth.}
\end{figure}

\clearpage

\begin{figure}
\epsscale{0.8} \plotone{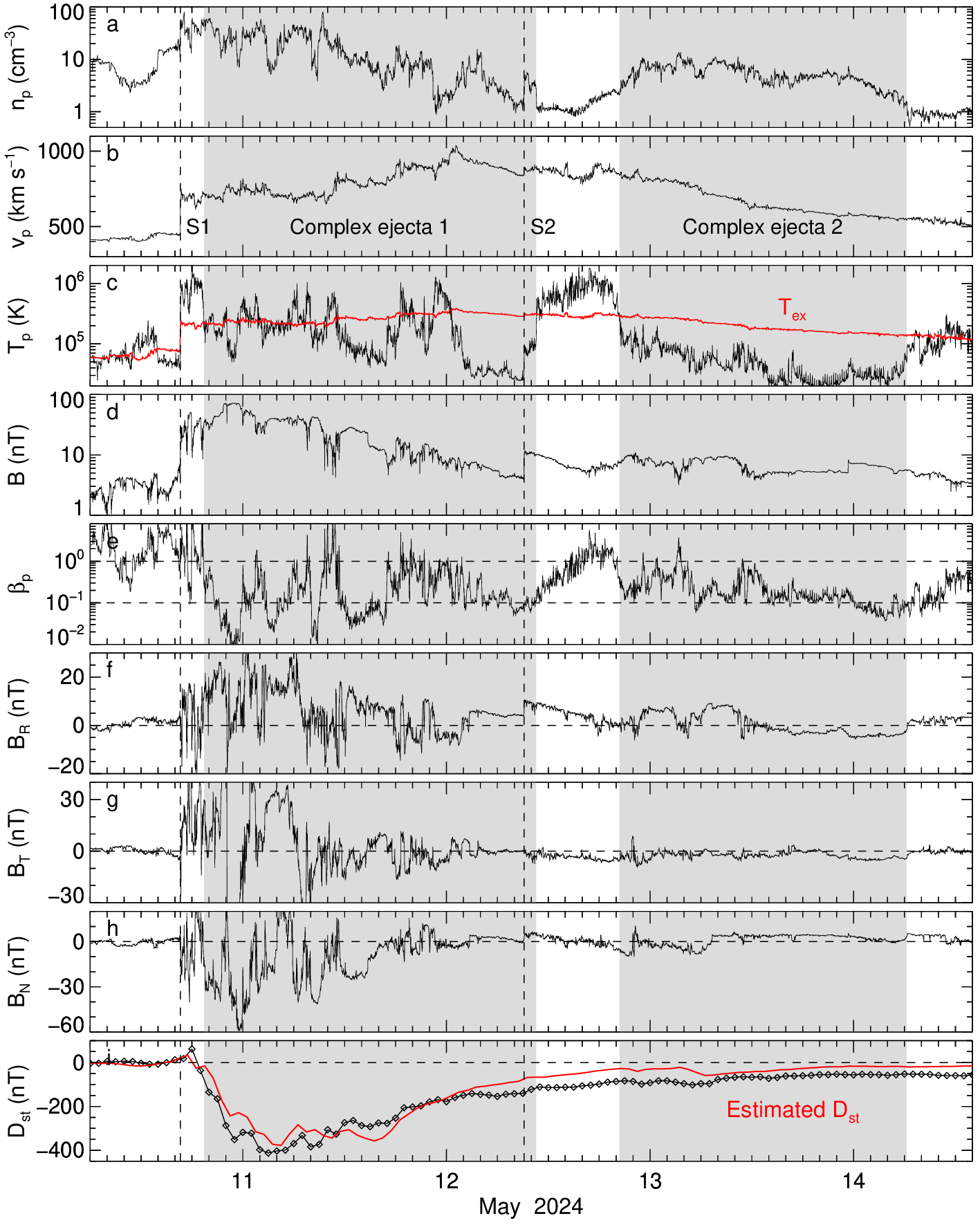} 
\caption{Solar wind parameters observed at Wind and measured $D_{\rm st}$ index. From top to bottom, the panels show the proton density (a), bulk speed (b), proton temperature (c), magnetic field strength (d), proton $\beta$ (e), magnetic field components (f-h), and $D_{\rm st}$ index (i), respectively. The shaded regions indicate the intervals of two complex ejecta, and the vertical dashed lines mark the associated shocks. The red curve in panel (c) denotes the expected proton temperature calculated from the observed speed \citep{lopez87, richardson95}. The red curve in panel (i) represents the estimated $D_{\rm st}$ index by combining the formulas of \citet{burton75} and \citet{om00}.}
\end{figure}

\clearpage

\begin{figure}
\epsscale{0.8} \plotone{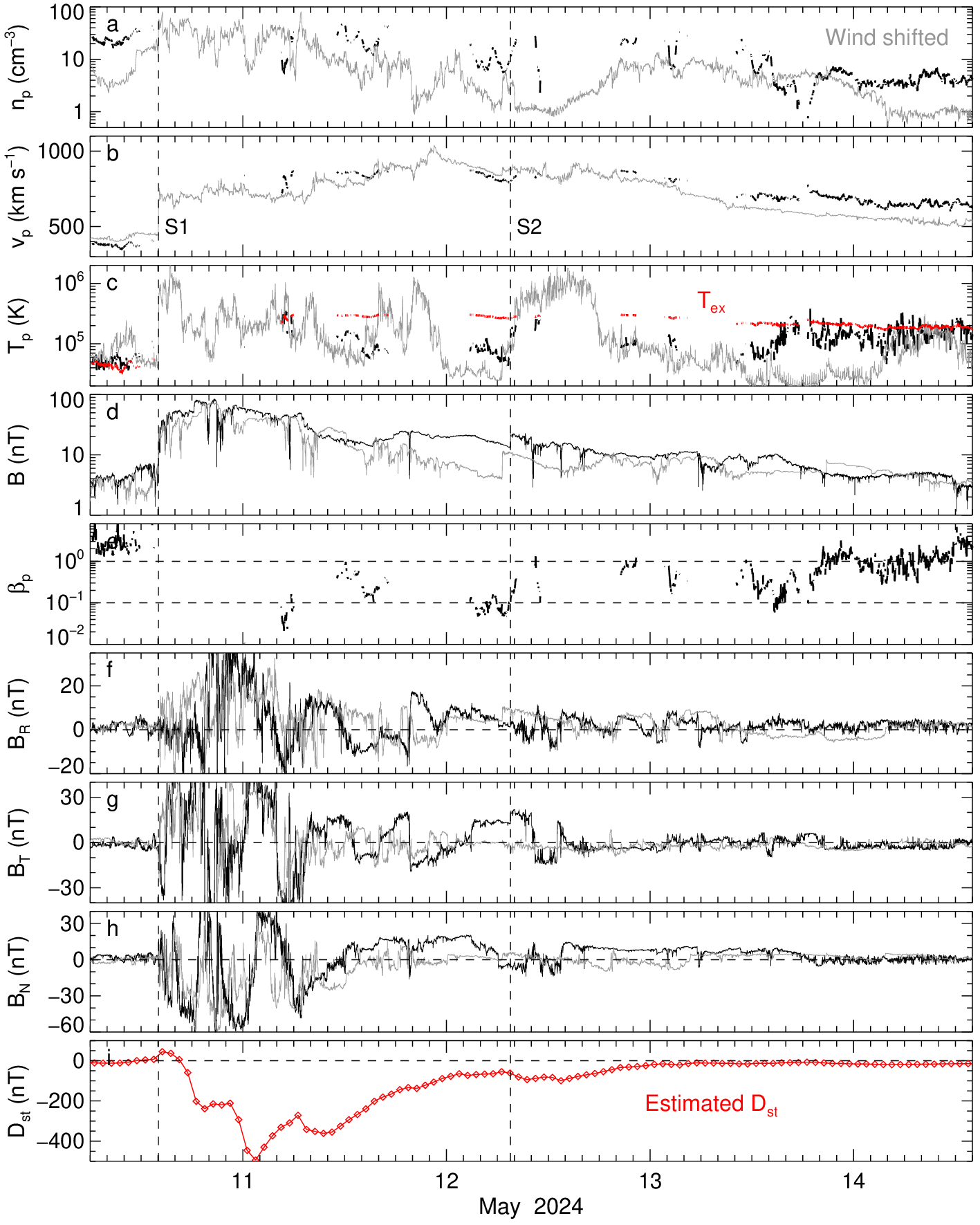} 
\caption{Solar wind parameters observed at STEREO A and estimated $D_{\rm st}$ index. Data at Wind, which are shifted forward in time by about 2.6 hr, are shown in gray for comparison. Similar to Figure~3.}
\end{figure}

\end{document}